\documentclass[sn-mathphys-num]{sn-jnl}

\usepackage{graphicx}%
\usepackage{epsfig,epstopdf}
\usepackage{multirow}%
\usepackage{amsmath,amssymb,amsfonts}%
\usepackage{amsthm}%
\usepackage{mathrsfs}%
\usepackage[title]{appendix}%
\usepackage{xcolor}%
\usepackage{textcomp}%
\usepackage{manyfoot}%
\usepackage{booktabs}%
\usepackage{algorithm}%
\usepackage{algcompatible}
\usepackage{algpseudocode}%
\usepackage{listings}%
\usepackage{tabularx}

\newcommand{\zerovec}{{\bf{0}}}

\newcommand{\Bb}{{\textbf{B}}}

\newcommand{\Db}{{\textbf{D}}}

\newcommand{\ub}{{\textbf{u}}}

\newcommand{\cb}{{\textbf{c}}}
\newcommand{\ab}{{\textbf{a}}}

\newcommand{\bb}{{\textbf{b}}}
\newcommand{\gb}{{\textbf{g}}}

\newcommand{\Ab}{{\textbf{A}}}

\newcommand{\rb}{{\textbf{r}}}

\newcommand{\vb}{{\textbf{v}}}

\newcommand{\Fb}{{\textbf{F}}}
\newcommand{\pb}{{\textbf{p}}}

\theoremstyle{thmstyleone}%
\theoremstyle{thmstyletwo}%

\theoremstyle{thmstylethree}%

\begin{document}

\title[Article Title]{Closed-form Single UAV-aided Emitter Localization and Trajectory Design Using Doppler and TOA Measurements}


\author[1]{\fnm{Samaneh} \sur{Motie}}\email{motie.s@qut.ac.ir}

\author*[1]{\fnm{Hadi} \sur{Zayyani}}\email{zayyani@qut.ac.ir}

\author[2]{\fnm{Mohammad} \sur{Salman}}\email{mohammad.salman@aum.edu.kw}

\author[3]{\fnm{Hasan} \sur{Abu Hilal}}\email{habuhilal@hct.ac.ae}


\affil[1]{\orgdiv{Department of Electrical and Computer Engineering}, \orgname{Qom University of Technology (QUT)}, \orgaddress{\street{Allahkaram street}, \city{Qom}, \country{Iran}}}

\affil[2]{\orgdiv{College of Engineering and Technology}, \orgname{American University of the Middle East}, \orgaddress{\city{Egaila 54200}, \country{Kuwait}}}

\affil[3]{\orgdiv{Electrical Engineering Department}, \orgname{Higher Colleges of Technology}, \orgaddress{\city{Abu Dhabi}, \country{UAE}}}


\abstract{In this paper, a single Unmanned-Aerial-Vehicle (UAV)-aided localization algorithm which uses both Doppler and Time of Arrival (ToA) measurements is presented. In contrast to Doppler-based localization algorithms which are based on non-convex functions, exploiting ToA measurements in a Least-Square (LS) Doppler-based cost function, leads to a quadratic convex function whose minimizer lies on a line. Utilizing the ToA measurements in addition to the linear equation of minimizer, a closed form solution is obtained for the emitter location using a constrained LS optimization. In addition, a trajectory design of the UAV is provided which has also closed-form solution. Simulation experiments demonstrate the effectiveness of the proposed algorithm in comparison to some others in the literature.}




\maketitle

\section{Introduction}
\label{sec:Intro}
Unmanned Aerial Vehicle (UAV) technology is a promising tool providing less expensive and more flexible solutions to applications such as communication \cite{Zhou19}, search and monitoring \cite{Yuan22}-\cite{He23}, search and rescue \cite{Atif21}-\cite{Kang23}, and localization \cite{Annep20}-\cite{Liu19}. UAV-assisted localization is a widespread application in the localization field \cite{Bayat23}-\cite{Liu19}. The localization task can be done by different features such as Received-Signal-Strength (RSS), Time-of-Arrival (ToA), Time-Difference-of-Arrival (TDoA), Angle-of-Arrival (AoA), Frequency-Difference-of-Arrival (FDoA), and Doppler \cite{Zeka19}.

Since the UAVs are moving, utilizing the FDoA and Doppler measurements is versatile in UAV-assisted localization related work \cite{Amar08}-\cite{Jiang19}. Among these, some uses Doppler only measurements \cite{Amar08}-\cite{Nguy18}, FDoA only measurements \cite{Pei22}, FDoA-TDoA \cite{Wang17}-\cite{Wang19}, Doppler-AoA \cite{Yin17}, and Doppler-ToA \cite{Jiang19}.

On the other classification approach of the above mentioned UAV-assisted localization related works, the most are utilizing multiple of UAVs for localization \cite{Bayat23}-\cite{Liu19}. Others, are localizing multiple emitters or targets such as \cite{Bayat23}-\cite{Liang22}, \cite{Jiang21}-\cite{Guer22}. Localizing simultaneously multiple emitters using only a single UAV is not possible or at least impossible until now. Moreover, localizing multiple emitters by utilizing multiple UAVs are possible \cite{Bayat23}-\cite{Liang22}. But, in order to reduce the cost, we utilize only a single UAV to find, of course, a single emitter. The main aim of this letter is to devise a single UAV-aided localization algorithm of a single emitter. There are some single UAV-aided localization algorithms in literature \cite{Yuan22}, \cite{Kang23}, \cite{Esra21}. Such methods have no closed-form solutions. On the other hand, many Doppler-based localization algorithms have a non-convex cost functions in their developed algorithms which hinder the efficiency of their proposed algorithms.

Following the comparative statements beforehand, in this letter, we propose a Doppler-based single UAV-aided localization algorithm. To remedy non-convexity nature of the suggested Least-Squares (LS) cost function, we use ToA measurements in addition to Doppler measurements. To best of our knowledge, there is only one mixed Doppler-ToA localization algorithm in the literature \cite{Jiang19} which differs from our proposed algorithm. Thanks to using ToA in addition to the Doppler, the LS cost function renders to a quadratic form which has not a unique solution and its minimizer lies on a line. Applying this linear condition along with ToA measurements results in a constrained optimization problem, for which the Lagrangian solution is derived in closed form. Moreover, a trajectory design of the UAV is also proposed to maintain the optimality of LS cost function which also has a closed-form solution. Simulation results also confirms the effectiveness of the proposed ToA-Doppler localization and trajectory design algorithm.

The rest of the paper is organized as follows. Section~\ref{sec:ProblemForm} discusses the system model and problem formulation. In Section~\ref{sec: prop}, the proposed localization algorithm is derived and developed, while the trajectory design is presented in Section~\ref{sec: Traj}. Simulation results are explained in Section~\ref{sec: Simulation}, while conclusions are drawn in Section~\ref{sec: con}.

\section{System Model and Problem Formulation}
\label{sec:ProblemForm}
\begin{figure}[tb]
\begin{center}
\includegraphics[width=7cm]{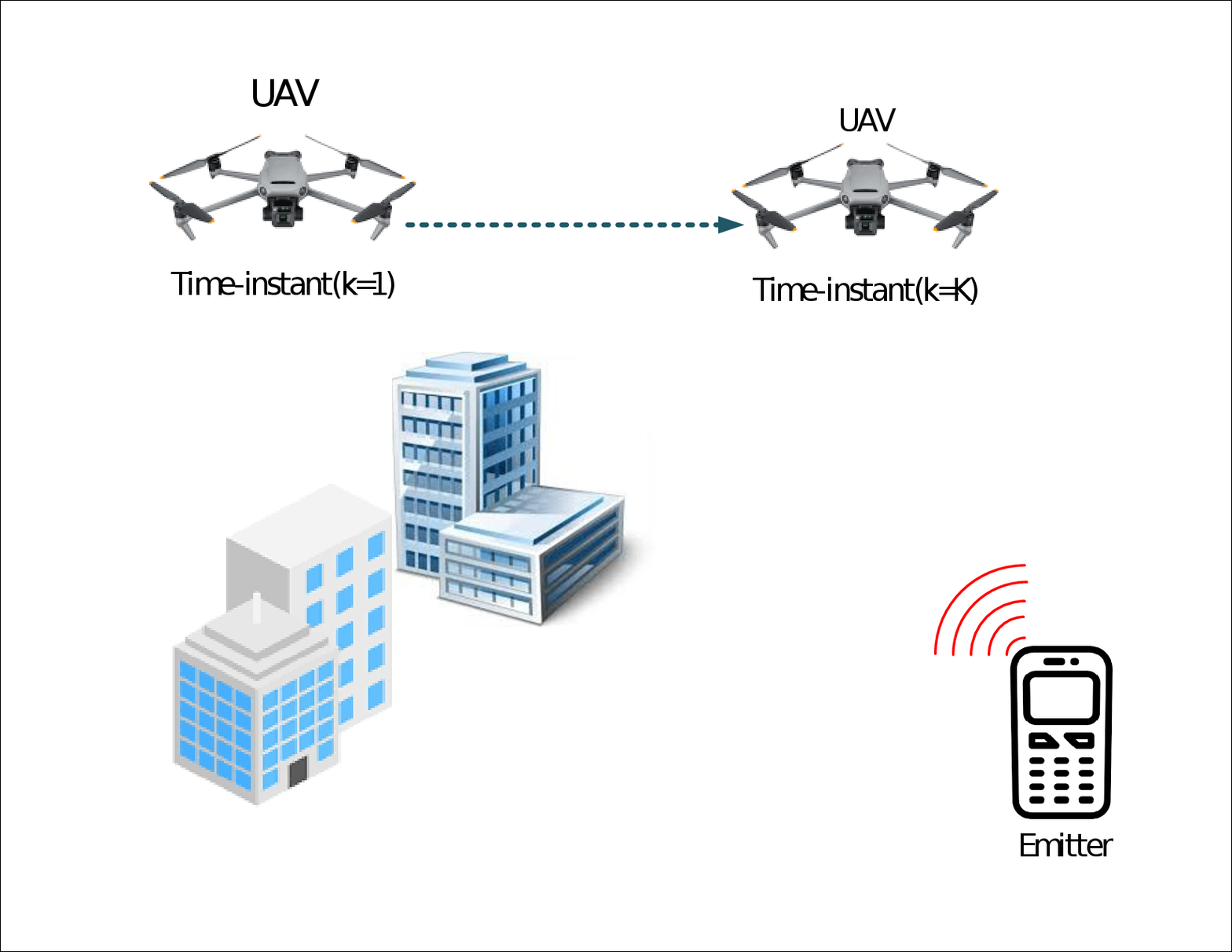}
\end{center}
\vspace{0.4cm}
\caption{The scenario of the problem which consists of a single moving UAV, and an emitter.}
\label{fig1}
\end{figure}

Assume a single UAV monitoring a region seeking for a single emitter (source) sending a Radio Frequency (RF) signal. The goal of single UAV is to estimate the location of the emitter which is denoted as $\pb_s=[x_s,y_s]^T$. The emitter is quasi-static which means that its displacement at the observation time is very small which can be neglected and can be considered as a fixed emitter. The scenario is depicted in Fig.~\ref{fig1}. The UAV position and velocity at time instant $k$ are $\pb_{u,k}=[x_{u,k},y_{u,k},z_{u,k}]^T$ and $\vb_{u,k}=[v_{x,u,k},v_{y,u,k},v_{z,u,k}]^T$ which are known. Each time instant is separated by a time step of $\Delta$. So, the time instant is $t_k=k\Delta$. The emitter emits a single tone signal with frequency $f_0$ which is received by the UAV. The UAV measures the Doppler frequencies due to relative motion between emitter and UAV. The Doppler frequency of the UAV at $k$'th time instant is equal to:

\begin{equation}
\small
f_{d,k}=\gamma\frac{\vb^T_{lk}(\pb_{u,k}-\pb_s)}{||\pb_{u,k}-\pb_s||}+w_{k},\quad 1\le k\le K,
\label{eq: fd}
\end{equation}

\noindent where $\gamma=\frac{f_0}{c}$ is a constant term, $K$ is the total number of measurements in each frame of measurements, and $w_{k}$ is the Doppler measurement noise (error) which is assumed to be zero-mean and with variance equal to $\sigma^2_w$. As it will be seen, to make the LS estimation problem convex, it is needed to have ToA measurements. Thus, ToA measurements are assumed available as

\begin{align}
\label{eq: TOA}
\tau_k=\frac{||\pb_{u,k}-\pb_s||}{c}+n_{\tau,k}, \quad 1\le k\le K.
\end{align}

\noindent where $n_{\tau,k}$ is the measurement error of ToA which is assumed to be zero-mean and with variance equal to $\sigma^2_{\tau}$. Now, the problem is to estimate the location of the emitter $\pb_s$ based on measuring all the Doppler frequencies of $f_{d,k}, \quad 1\le k\le K$ and all the ToA measurements of $\tau_{k}, \quad 1\le k\le K$ assuming that it is quasi-fixed in a single frame of measurements. Thus, after the rough estimation of emitter location,  new trajectory for the UAV, based on minimizing the LS cost function, is designed and presented in the next section. Thus, at the end of each frame, the location estimation and the new trajectory of the next frame is designed.

\section{The Proposed Localization Algorithm}
\label{sec: prop}
Let us assume that in the first frame of measurements ($1\le k\le K$), the velocity of UAV is fixed ($\vb_{u,k}=\vb_u$) and the path of UAV is straight. The main idea of using a single UAV is to gather all Doppler measurements of UAV in different positions to be able to estimate the location of emitter. So, we have $\pb_{u,k}=\pb_{u,1}+(k-1)\Delta\vb_u$. Based on (\ref{eq: fd}), we have
\begin{align}
f_k||\pb_{u,k}-\pb_s||=\gamma\vb^T_k(\pb_{u,k}-\pb_s)+w_k||\pb_{u,k}-\pb_s||.
\end{align}
So, the LS cost function will be
\begin{align}
\label{eq: gls}
\mathrm{g}_{\mathrm{LS}}(\pb_s)=\sum_{k=1}^K\Big[f_k||\pb_{u,k}-\pb_s||-\gamma\vb^T_k(\pb_{u,k}-\pb_s)\Big]^2.
\end{align}
As it can be seen from (\ref{eq: gls}), this LS cost function is generally non-convex. But, using the ToA measurements in (\ref{eq: TOA}), and replacing into (\ref{eq: gls}), we have the following modified quadratic convex LS cost function:
\begin{align}
\label{eq: glsf}
\mathrm{\bar{g}}_{\mathrm{LS}}(\pb_s)=\sum_{k=1}^K\Big[\bar{f}_k-\gamma\vb^T_k(\pb_{u,k}-\pb_s)\Big]^2,
\end{align}
where $\bar{f}_k\triangleq c\tau_kf_k$. Since, the modified LS cost function is quadratic, it has a closed-form minimizer. To find the solution, we make the partial derivative and enforce it to be zero vector. So, we have:
\begin{align}
\label{eq: der}
\frac{\partial \mathrm{\bar{g}}_{\mathrm{LS}}(\pb_s)}{\partial\pb_s}=\sum_{k=1}^K-2\gamma\Big[\bar{f}_k-\gamma\vb^T_k(\pb_{u,k}-\pb_s)\Big]\vb_k=\zerovec.
\end{align}
Hence, (\ref{eq: der}) leads to the following linear equation
\begin{align}
\label{eq: rr}
\gamma\sum_{k=1}^K\vb^T_k(\pb_{u,k}-\pb_s)\vb_k=\cb,
\end{align}
where $\cb=\sum_{k=1}^K\bar{f}_k\vb_k$ is a known vector. By some simple manipulations to (\ref{eq: rr}), we have
\begin{align}
\label{eq: rr1}
\sum_{k=1}^K\vb^T_k\pb_s\vb_k\vb_k=\cb_1,
\end{align}
where $\cb_1\triangleq -\frac{1}{\gamma}\cb+\sum_{k=1}^K\vb^T_k\pb_{u,k}\vb_k$ is a known vector. So, assuming that the UAV has a fixed height equal to $z_{u,k}=h$ and the UAV's fixed velocity at a single frame is $\vb_k=\vb=[v_{k,1},v_{k,2},0]^T$, from (\ref{eq: rr1}) we have
\begin{align}
\label{eq: rr2}
\sum_{k=1}^K(v_{k,1}x_s+v_{k,2}y_s)\vb^{-}_k=x_s\ab_1+y_s\ab_2=\cb^{-}_1,
\end{align}
where $\vb^{-}_k=[v_{k,1},v_{k,2}]^T$, $\cb^{-}_1=[\cb_{11},\cb_{12}]^T$, $\ab_1\triangleq \sum_{k=1}^Kv_{k,1}\vb^{-}_k$, and $\ab_2\triangleq \sum_{k=1}^Kv_{k,2}\vb^{-}_k$. Finally, (\ref{eq: rr2}) can be written in the following linear system of equations:
\begin{align}
\Ab\pb_s=[\ab_1|\ab_2]\pb_s=\cb^{-}_1,
\end{align}
where $\Ab=[\ab_1|\ab_2]$ is a $2\times 2$ matrix. Since it is assumed that in a frame of measurements ($K$ consecutive measurements), the velocity of UAV is constant equal to $\vb_k=\vb_0$, the determinant of $\Ab$ is zero since $\ab_1=\vb^{-}_0\sum_{k=1}^Kv_{k,1}$, and $\ab_2=\vb^{-}_0\sum_{k=1}^Kv_{k,2}$ are dependent. So, it is not possible to estimate the location of emitter just by minimizing the LS function. But, fortunately, minimizing the LS cost function yields a linear condition which would help to estimate the location of emitter as it will be explained next. The linear condition is
\begin{align}
\label{eq: lineq}
k_1x_s+k_2y_s=c_{11},
\end{align}
where $k_1\triangleq v_{01}\sum_{k=1}^Kv_{k,1}$ and $k_1\triangleq v_{01}\sum_{k=1}^Kv_{k,2}$. In fact, the minimizer of LS cost function lies on this line. In fact, it is not possible to find the location of emitter just by measuring Doppler measurements using a single UAV. Now, we use the ToA measurements in addition to Doppler measurements. The ToA measurements are
\begin{align}
||\pb_{u,k}-\pb_s||^2=d^2_k.
\end{align}
Similar to other ToA localization algorithms \cite{Zeka19}, the ToA measurements can be written in the following form:
\begin{align}
\label{eq: Beq}
\Bb\rb=\pb,
\end{align}
where
$p_k=d^2_k-x^2_{u,k}-y^2_{u,k}$, $\rb=[r_x,r_y,r_z]^T=[x_s,y_s,x^2_s+y^2_s]^T$, and we have
\begin{align}
\Bb=\left(
  \begin{array}{ccc}
    -2x_{u,1} & -2y_{u,1} & +1 \\
    -2x_{u,2} & -2y_{u,2} & +1 \\
    \vdots & \vdots & \vdots \\
    -2x_{u,K} & -2y_{u,K} & +1 \\
  \end{array}.
\right)
\end{align}
Combining (\ref{eq: Beq}) and (\ref{eq: lineq}), we have
\begin{align}
\Bb^{+}\rb=\pb^{+},
\end{align}
where $\Bb^{+}$ is a $(K+1)\times 3$ matrix which has an extra row in comparison to $\Bb$ equal to $\Bb(K+1,:)=[k_1,k_2,0]$, and $\pb^{+}$ is a $(K+1)\times 1$ vector which has an extra element in comparison to $\pb$ equal to $p^{+}_{K+1}=c_{11}$. The vector $\rb$ lies in the equation of $r_z=r^2_x+r^2_y$ which the matrix form is
\begin{align}
\rb^T\Db\rb+2\gb^T\rb=0,
\end{align}
where $\gb=[0,0,-\frac{1}{2}]^T$ and
\begin{align}
\Db=\left(
      \begin{array}{ccc}
        1 & 0 & 0 \\
        0 & 1 & 0 \\
        0 & 0 & 0 \\
      \end{array}.
    \right)
\end{align}
Hence, the location estimation problem is indeed a constrained LS (CLS) problem in which the Lagrange multiplier cost function is equal to
\begin{align}
\mathrm{J}(\rb)=||\Bb^{+}\rb-\pb^{+}||^2+\lambda(\rb^T\Db\rb+2\gb^T\rb),
\end{align}
where $\lambda$ is the Lagrange multiplier. The minimizer of the cost function of CLS is easily seen to be
\begin{align}
\label{eq: finalest}
\rb^{*}=\Fb^{-1}(\Bb^{+}\pb^{+}-\lambda\gb),
\end{align}
where $\Fb=\Bb^{+T}\Bb^{+}+\lambda\Db$.


\section{Trajectory design of the single UAV}
\label{sec: Traj}
In this section, the trajectory design of the single UAV is presented. After each frame of measurements (for example after the first frame), the location estimation of stationary emitter is performed as presented in section~\ref{sec: prop}. Then, at the beginning of the next frame (for example the second frame), we modify the trajectory of the UAV based on minimizing the LS cost function. For minimum latency, we used just the first Doppler measurement at the next frame. The augmented LS cost function based on adding just one measurements of the next frame is equal to
\begin{align}
\mathrm{\bar{g}}^{+}_{\mathrm{LS}}(\pb_s)=\sum_{k=1}^{K+1}\Big[\bar{f}_k-\gamma\vb^T_k(\pb_{u,k}-\pb_s)\Big]^2.
\end{align}
Now, since the location is estimated at previous frame, we can replace $\hat{\pb}_s$ instead of $\pb_s$ and write the augmented LS cost function as
\begin{align}
\label{eq: aLS}
\mathrm{\bar{g}}^{+}_{\mathrm{LS}}(\pb_s)\approx\mathrm{\bar{g}}_{\mathrm{LS}}(\pb_s)+\Big[\bar{f}_{K+1}-\gamma\vb^T_{K+1}(\pb_{u,K+1}-\hat{\pb}_s)\Big]^2.
\end{align}
Now, minimizing the augmented LS cost function in (\ref{eq: aLS}) with respect to $\vb_{K+1}$ is equivalent to minimizing the following cost function as
\begin{align}
\label{eq: gg}
\mathrm{g}(\vb)=\Big[\bar{f}_{K+1}-\gamma\vb^T(\pb_{K}+\Delta\vb-\hat{\pb}_s)\Big]^2,
\end{align}
where $\vb=\vb_{K+1}$ is the unknown new velocity of the UAV, and $\pb_{K}=\pb_{u,K}$ is the known position of UAV at the end of first frame. To minimize the cost function of (\ref{eq: gg}), we take the derivative and enforce it to be zero vector. So, we have
\begin{align}
\frac{\partial\mathrm{g}(\vb)}{\partial\vb}=
-2\gamma\Big[\bar{f}_{K+1}-\gamma\vb^T(\pb_{K}+\Delta\vb-\hat{\pb}_s)\Big](\pb_K-\hat{\pb}_s+2\Delta\vb)=\zerovec.
\label{eq: gg1}
\end{align}
Since $\pb_K-\hat{\pb}_s+2\Delta\vb\neq\zerovec$, the condition of (\ref{eq: gg1}) leads to
\begin{align}
\label{eq: cond1}
\bar{f}_{K+1}=\gamma\vb^T(\pb_{K}+\Delta\vb-\hat{\pb}_s).
\end{align}
If we nominate $\vb=[v_1,v_2,0]^T$, $\vb^{-}=[v_{1},v_{2}]^T$, $\pb_{K}=[p_1,p_2,p_3]^T$, and $\pb^{-}_{K}=[p_1,p_2]^T$, then, (\ref{eq: cond1}) can be written in the following form:
\begin{align}
\label{eq: cond1}
\bar{f}_{K+1}=\gamma\vb^{-T}\pb^{-}_{K}+\gamma\Delta\vb^{-T}\vb^{-}-\gamma\vb^{-T}\hat{\pb}_s.
\end{align}
This, in term, leads to
\begin{align}
\label{eq: cond2}
\bar{f}_{K+1}-\gamma\Delta||\vb^{-}||^2=\gamma\vb^{-T}(\pb^{-}_{K}-\hat{\pb}_s).
\end{align}
If we denote $||\vb||^2=||\vb^{-}||^2=A^2_v$ as a constant term\footnote{It is assumed that the velocity magnitude is equal to $A_v$ and fixed otherwise stated.}, the condition of (\ref{eq: cond2}) leads to
\begin{align}
\label{eq: ueq}
\ub^T\bb=d,
\end{align}
where $\ub=\vb^{-}$ is the unknown new velocity of the UAV, $d=\frac{1}{\gamma}(\bar{f}_{K+1}-\gamma\Delta||\vb^{-}||^2)$ is a known scalar, and $\bb=\pb^{-}_{K}-\hat{\pb}_s$ is a known vector. To find $\ub=[u_x,u_y]^T$ from (\ref{eq: ueq}), we have the following nonlinear system of equations:
\begin{align}
\label{eq: systemeq}
\Bigg\{
    \begin{array}{ll}
    u_xb_1+u_yb_2=d,\\
    u^2_x+u^2_y=A^2_v.
    \end{array}
\end{align}
Therefore, solving (\ref{eq: systemeq}) in terms of $u_x$, leads to the following second order equation as
\begin{align}
\label{eq: ord2}
(1+b^2_1)u^2_x-2(\frac{db_1}{b^2_2})u_x+(\frac{d^2}{b^2_2}-A^2_v)=a^{'}u^2_x-2b^{'}u_x+c^{'}.
\end{align}
If $\Delta_2\triangleq b^{'2}-a^{'}c^{'}\ge 0$, the equation (\ref{eq: ord2}) have two solutions of $u_{x,1,2}=-\frac{b^{'}}{a^{'}}\pm\frac{\sqrt{\Delta_2}}{a^{'}}$. To select one solution among two solutions, we select the one that the next position of UAV is nearer to emitter than the other solution. But, if the $\Delta_2<0$, then the equation (\ref{eq: ord2}) has not solution. Since $a^{'}>0$, then the second order equation is non negative. In this case, to find a solution, we violate our first assumption of constant velocity magnitude and change the value of $A^2_v=||\vb||^2$. We use a value for velocity magnitude to have $\Delta_2=0$ to prevent the double solutions. So, some manipulations show that a genuine suggestion is to choose
\begin{align}
\label{eq: dnew}
A^2_{v,new}=\frac{d_{new}^2}{b^2_2(1+b^2_1)},
\end{align}
where $d_{new}=\frac{1}{\gamma}(\bar{f}_{K+1}-\gamma\Delta A^2_{v,new})$ is the new scalar based on new velocity magnitude. Replacing this new scalar into (\ref{eq: dnew}) and solving for $A^2_{v,new}$, with some manipulations, we have the final formula for finding $A^2_{v,new}$
\begin{align}
\label{eq: A2}
\gamma^2\Delta^2A^4-(2\gamma\Delta-\frac{1}{l})A^2+\bar{f}^2_{K+1}=a^{''}A^4+b^{''}A^2+c^{''}=0,
\end{align}
where $l\triangleq \frac{1}{\gamma^2}\frac{1}{b^2_2(1+b^2_1)}$. If $\Delta_4=b^{''2}-4a^{''}c^{''}\ge 0$, we have two positive solutions, or two negative solutions of $A_{v,new,1}$ and $A_{v,new,2}$ (since the product of solutions is equal to $\frac{c^{''}}{a^{''}}>0$). If two solutions are negative, the trajectory optimization have no solution and hence we did not perform the trajectory optimization. Otherwise, if two solutions are non negative, we select the minimum solution to not violating the condition of $A_v\le V_{max}$. So, the final velocity magnitude is equal to $A^{f}_{v,new}=\mathrm{min}\{\mathrm{min}\{A_{v,new,1},A_{v,new,2}\},V_{\mathrm{max}}\}$. Finding $A^f_{v,new}$ and then we can find $d^f_{new}$ from $d^f_{new}=\frac{1}{\gamma}(\bar{f}_{K+1}-\gamma\Delta A^{2,f}_{v,new})$. Then, the new solution for the velocity is
\begin{align}
\label{eq: ufin}
u^f_x=-\frac{b^{'}}{a^{'}}=\frac{\frac{d^f_{new}b_1}{b^2_2}}{1+b^2_1}.
\end{align}
Put all together, the overall algorithm is to estimate the location of emitter in closed-form based on (\ref{eq: finalest}) at the end of $l$'th frame with time measurements of $(l-1)K+1\le k\le lK$. Then, the new velocity which determines the trajectory at the $l+1$'th frame is obtained by solving (\ref{eq: ord2}) and selecting the nearer solution if $\Delta_2\ge0$, or otherwise by (\ref{eq: ufin}) following the approach explained before. Again, at the end of $l+1$'th frame, the new estimation of location of emitter is calculated the same as before and this continues. The outline of the proposed localization and trajectory design algorithm is outlined in Table 1.

\begin{algorithm}[t]

\caption{The proposed Doppler-ToA localization and trajectory design algorithm.}
\label{Algorithm_1}
\textbf{Input}   Doppler measurements: $f_{d,k}$, ToA measurements $\tau_k$, $f_0$. \newline
\textbf{Initialize} initial velocity of UAV: $\vb_0$, Initial position of UAV: $\pb_{u,0}$, $\ub_0=\vb_0$, and $l=1$.

\REPEAT Repeat\\

\STATE
 {1. Measure the Doppler measurements $f_{d,k}$ and ToA measurements $\tau_k$ for $(l-1)K+1\le k\le lK$};\\
\STATE
 {2. Estimate the Emitter location in closed-form based on (\ref{eq: finalest}});\\
\STATE
 {3. Measure the new Doppler measurement $f_{K+1}=f_{d,lK+1}$}\\
 \STATE
{4. Find the new velocity by fixed assumption of velocity magnitude and by solving (\ref{eq: ord2}) in closed-form and selecting the nearer solution if $\Delta_2\ge0$, Otherwise change the velocity magnitude and find the new velocity from (\ref{eq: ufin}) following the approach explained in Section~\ref{sec: Traj}};\\
\STATE 5. $l \leftarrow l+1$;\\
\UNTIL { Until a stopping criterion is reached.\\}
\textbf{Output} Location of emitter $\hat{\pb}_s$ at the end of $l$'th frame and constant velocity of UAV at the beginning of $l$'th frame;
\end{algorithm}



\section{Simulation Results}
\label{sec: Simulation}
\begin{figure}[ht]
\begin{center}
\includegraphics[width=7.5cm]{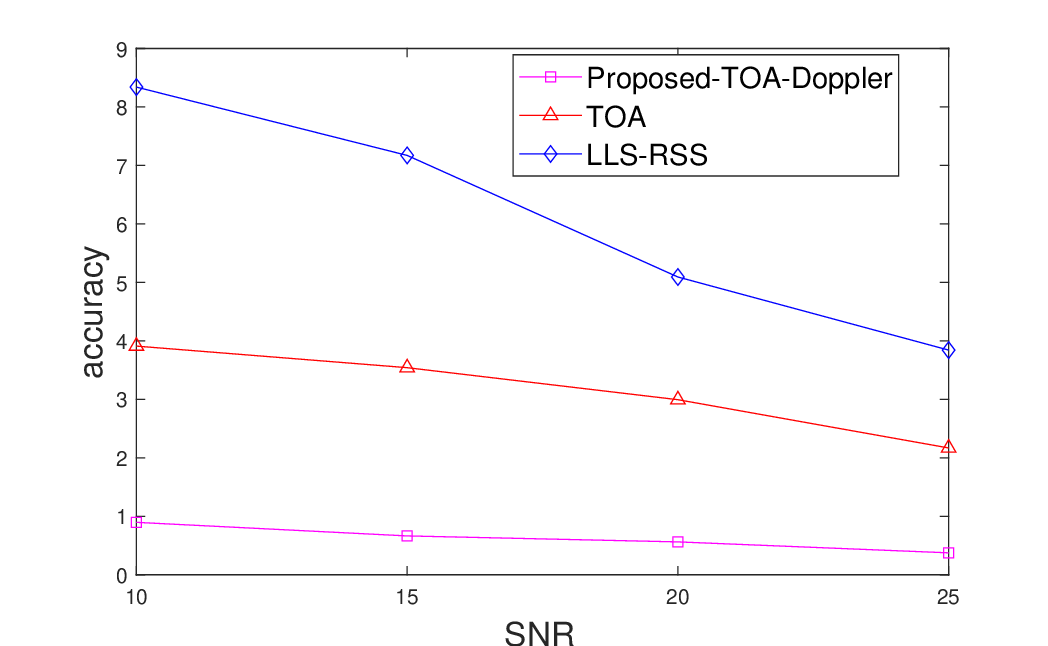}
\end{center}
\vspace{-0.5 cm}
\caption{Position-accuracy versus SNR.}
\label{fig2}
\end{figure}

In this section, the performance of the proposed UAV-aided location estimation of the emitter is investigated.
The experiment is simulated using a single UAV which the initial position of the UAV is selected as $[0,0,50]$ and the z-coordinate od the UAV is always selected as $z=50$ which means that UAV has a fixed height. The UAV at each frame has a fixed velocity which is $\vb_0=[10,0,0]^T$. The position of emitter is assumed to be fixed as $\pb_0=[35,15,0]^T$. The frequency of single tone emitter of the vehicle is equal to $f_0=3\times10^8$. The time step is equal to $\Delta=0.05$, and the number of measurements in a frame is $K=10$. The variances of Doppler noise and ToA measurement noise are assumed to be $\sigma^2_d=0.01$, and $\sigma^2_{\tau}=1e-6$, respectively, unless otherwise stated. The performance metric for location estimation is the distance accuracy which is concisely nominated as accuracy and defined as $\mathrm{Accuracy}=||\pb_0-\hat{\pb}_0||$, where $\pb_0$ is the initial position of the emitter, and $\hat{\pb}_0=[\hat{\pb}_1,\hat{\pb}_2,\hat{\pb}_3]^T$ is the estimated position of the emitter. The reported accuracy of position is the average result of 50 independent random Monte Carlo simulations.
In the first experiment, the position estimation of the emitter without trajectory optimization is investigated. The simulation parameters are stated before. The competing algorithms are LLS-RSS \cite{So11}, and the conventional ToA algorithm \cite{Zeka19}. The SNR of each algorithm which is SNR of RSS, SNR of ToA, and SNR of Doppler, is varied and the accuracy is calculated in each case. The accuracy versus SNR is depicted in Figure~\ref{fig2}. It shows that the accuracy is improved (accuracy value is decreased) by increasing the SNR and the best algorithm is the proposed ToA-Doppler method which used both features for localization. It is worth mentioning that it is not possible to localize the emitter using only Doppler measurements of only one single UAV.

\vspace{-0.5cm}

\begin{figure}[ht]
\begin{center}
\includegraphics[width=7.5cm]{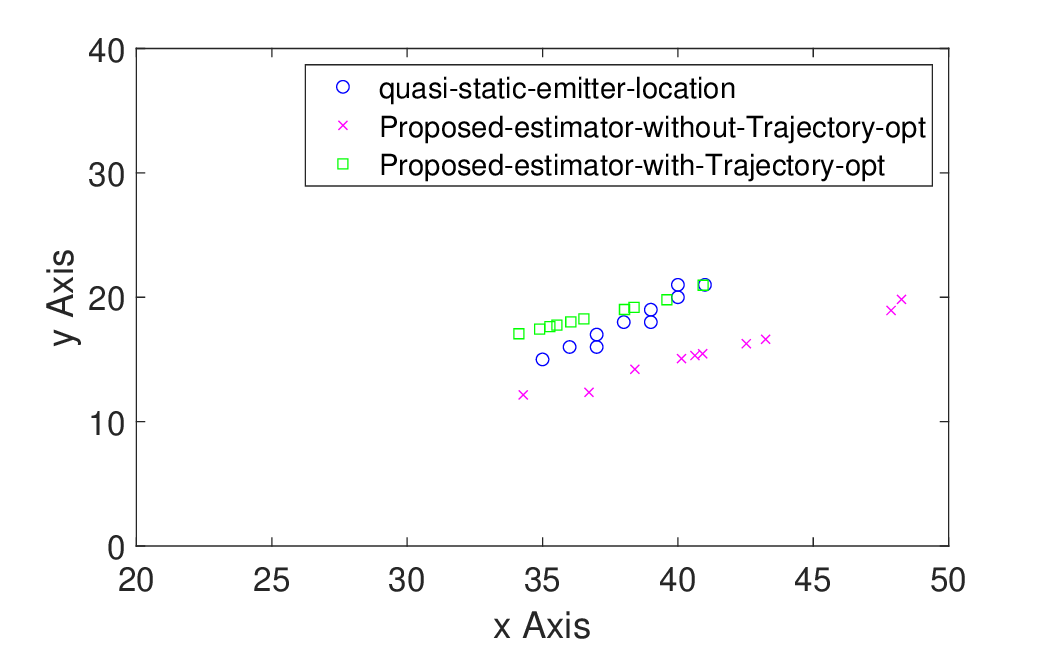}
\end{center}
\vspace{-0.5 cm}
\caption{The scenario of trajectory optimization.}
\label{fig3}
\end{figure}

In the second experiment, the effectiveness of the trajectory optimization or design is investigated. Again, the parameters are the same as stated before. Now, we use 10 frames each with $K=10$ measurements of Doppler and ToA. The emitter has a quasi-static position and the UAV has a constant optimized velocity or linear trajectory at each frame. To compare the performance, two cases are considered. At first case, the UAV has its initial velocity in all frames and do not use trajectory optimization. At the second case, the velocity of the UAV is optimized at the beginning of each frame as outlined in Algorithm 1. The position of quasi-static emitter, and the location position estimator of emitter in the two cases, are depicted in Figure~\ref{fig3}. It shows that the proposed localization method with trajectory optimization can estimate the location of emitter effectively, while the proposed method without trajectory optimization get worse results after several frames since in this case, the UAV goes far away from the emitter and it could not correct its trajectory.

%

\section{Conclusions and Future Work}
\label{sec: con}
In this letter, a joint localization and trajectory optimization algorithm of a single UAV-aided localization scenario with a quasi-stationary emitter is presented. In the presented algorithm, both Doppler and ToA measurements are utilized. Aiding ToA measurements in a Doppler-based LS cost function results in a quadratic convex cost function. However, the Linear equation of the minimizer is redundant and hence the minimized lies on a line. Utilizing this linear condition and ToA measurements leads to an optimization problem in which its Lagrangian solution has a closed form. To monitor the emitter and improve its localization, a trajectory optimization design is developed based on the augmented Least-Squares (LS) cost function, which also has a closed-form solution. The simulation results showed the advantages of the proposed algorithm in localizing a quasi-static emitter such as a pedestrian. Also, it has a potential application in guiding and localizing blind persons using a single UAV which will be for future works.


\end{document}